\documentclass[twocolumn,aps,prl,showpacs, showkeys,groupedaddress,amsmath,amssymb]{revtex4}
\usepackage{amsmath}
\usepackage{amssymb}
\usepackage{txfonts}
\usepackage{graphicx}
\begin{document}
\title{Quantum Communications with Compressed Decoherence Using
Bright Squeezed Light}
\author{Fang-Yu Hong}
\author{Shi-Jie Xiong}

\affiliation{National Laboratory of Solid State Microstructures and
Department of Physics, Nanjing University, Nanjing 210093, China}
\date{\today}
\begin{abstract}
We propose a scheme for long-distance distribution of quantum
entanglement in which the entanglement between qubits at
intermediate stations of the channel is established by using bright
light pulses in squeezed states coupled to the qubits in cavities
with a weak dispersive interaction.  The fidelity of the
entanglement between qubits at the neighbor stations (10 km apart
from each other) obtained by postselection through the balanced
homodyne detection of 7 dB squeezed pulses  can reach $F=0.99$
without using entanglement purification, at same time, the
probability of successful generation of entanglement is 0.34.
\end{abstract}

 \maketitle

Quantum communication holds promise for transmitting secure messages
via quantum cryptography and for distributing quantum information
\cite{Giulio}, and is an essential element in the construction of
quantum networks. The basic problem of quantum communication is to
create nearly perfect entangled states between remote stations. Such
entanglement can be used, for example, to perform an entanglement
swap \cite{dur,bri}, to faithfully transfer quantum states via
quantum teleportation, and to implement secure quantum cryptography
using the Ekert protocol \cite{ekert}. At present, photonic channels
are the only choice in the realistic scheme for quantum
communication. However, the fidelity of entanglement between remote
sites decreases exponentially with the length of the connecting
channel because of optical absorption and other channel noise, which
limits the range of direct quantum communication techniques
\cite{bra}. To extend this range to longer distance remains a
conceptual and technological challenge.

In principal, the exponential fidelity decay can be overcome by
introducing intermediate quantum nodes and utilizing the so-called
quantum repeater protocol \cite{bri}. In a repeater protocol, an
entanglement is established over long distances by building a
backbone of entangled pairs between closely spaced nodes. Performing
an entanglement swap at each intermediate node leaves the outer two
nodes entangled. Even though quantum operations are subject to
errors, by incorporating entanglement purification \cite{ben,deu} at
each step, in principle, the overall communication fidelity can be
made very close to unity, with the communication time growing only
polynomially with the transmission distance.

Nearly all of the existing schemes
\cite{cir,yao,duan,chi,chil,wak,duanl,kraus} for quantum repeaters
generating entanglement rely on single photon (or sub-photon
coherence state) transmission  between distant qubits. Although the
heralded entanglement may have high initial fidelity, the high
probability for the channel loss means that the probability of
successful entanglement decreases exponentially with the distance
between repeater stations and the communication rates will be very
low. Loock {\it et al} \cite{loo,lad} have described a scheme which
uses a bright coherence pulse instead of single photons as the
mediator linking the remote nodes and has successful post-selection
of about 36\% of the pulse sent down the channel with initial
fidelity 0.77 in the post-selected entanglement.

Schemes for generation of entanglement between two qubits through
squeezed lights have been suggested \cite{wsmk,mpws,bkjc}. In this
paper we describe a scheme for distributing  entanglement based on
dispersive light-matter interactions using bright 7 dB squeezed
light pulses which have already been experimentally realized
\cite{pklr}. Squeezed lights with larger squeeze factor may be
achieved through the approach proposed by Guzm\'{a}n {\it et al}
\cite{guz}. The fidelity of entanglement between the two qubits in
neighbor nodes may be as high as 0.99, at the same time, the
probability of successful postselection can reach
 0.34. So this scheme may significantly
enhance the quantum communication rates.

In general, a nonlinear element has to be introduced to implement
long-distance quantum communication \cite{loo} in at least two
possible ways. The first approach employs only linear
transformations accompanied with a measurement-induced nonlinearity
\cite{kni}. In the second method, a weak nonlinear gate, where the
nonlinearity is efficiently enhanced with a sufficiently strong
probe pulse, is used to supplement the linear gates \cite{nem}. This
concept can be applied to quantum communication with a hybrid system
based on optical-carrier waves and electron-spin signals. In our
scheme, a bright probe pulse in squeezed state sequentially
interacts with two electronic spins in cavities at neighboring
repeater stations, so the postselection of entangled qubit pairs
will be conditional upon the results of probe homodyne measurements.
In our proposal the complication of purifying an atomic ensemble
\cite{duan} can be avoided and the entanglement from several copies
of noisy entangled pairs of electronic spins may be directly
distilled.

A qubit may be naturally realized through electron spins such as
single electrons trapped in quantum dots \cite{abra} and  neutral
donor impurities in semiconductors \cite{str,fu}. The system should
be placed in a high-Q cavity resonant with the light to have a
sufficiently weak dispersive interaction between the electron and
the light. For the cavity, a weak coupling is sufficient \cite{lad,
hon}.

\begin{figure}
\includegraphics[scale=0.35]{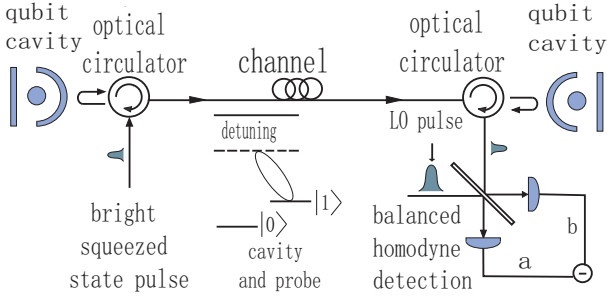}
\caption{\label{fig:1}( Color online) Schematic diagram showing the
generation of spin entanglement between two qubits at neighboring
nodes via homodyne detection discriminating among conditionally
phased-rotated squeezed states of probe pulses. The local (LO) pulse
for the homodyne detection is in a coherence state. }
\end{figure}

The mechanism for the entanglement distribution among the
neighboring stations in the channel is shown in Fig. \ref{fig:1}.
The system in the cavity is treated as a $\Lambda$ system, with two
stable or metastable ground states $|0\rangle$ and $|1\rangle$, and
only $|1\rangle$ takes part in the interaction with the cavity mode.
Local rotations between $|0\rangle$ and $|1\rangle$ may be
manipulated via stimulated Raman transitions. Particularly, we
assume that the states of the qubits in the neighboring nodes are
both in the state $(|0\rangle+|1\rangle)/\sqrt{2}$. The probe pulse
is sufficiently detuned from the transition between $|1\rangle$ and
the exited state to guaranty a strictly weak dispersive light-matter
interaction.

The probe beam is in a squeezed state $|\alpha,\varepsilon\rangle$
defined as \cite{wal}
\begin{equation}
|\alpha,\varepsilon\rangle=e^{i\alpha a^\dagger-\alpha^\ast
a}e^{\frac{1}{2}\varepsilon^\ast a^2-\frac{1}{2}\varepsilon
a^{\dagger2}}|0\rangle,
\end{equation}
where $a$ is  the annihilation operator of a quantum field mode,
$\alpha$ is an arbitrary  complex number, and
$\varepsilon=re^{2i\phi}$ with $r$ being the squeeze factor.

We assume that $\alpha$ is real and $\varepsilon=r e^{\,i\pi}$
hereafter. When the probe beam in the squeezed state
$|\alpha,\varepsilon\rangle$ reflects from the cavity, the total
output state may be described by \cite{hon}
\begin{equation}
\hat{U}_{int}[(|0\rangle+|1\rangle)|\alpha,\varepsilon\rangle]/\sqrt{2}=
(|0\rangle|\alpha,\varepsilon\rangle+|1\rangle|\alpha
e^{-i\theta},\varepsilon e^{-i2\theta}\rangle)/\sqrt{2}.
\end{equation}
For semiconductor impurities and realistic cavity parameters, phase
shifts of $\theta$ about $0.01$ are achievable \cite{loo,hon}. The
change in $\varepsilon$ and the amplitude of $\alpha$ are extremely
small and may be omitted because of the weak dispersive interaction
between light and matter.

In the physical optic fiber, photon losses are inevitable, which
results in the squeezed pulse to degrade. So in a long distance
optic fiber  communication with large losses, the transmittance
$\eta^2\ll1$, there will be no advantage in communicating via
squeezed states. Otherwise, a part of the noise reduction of the
pulse can be conserved, and the pulse may be assumed in a squeezed
state $|\alpha',\varepsilon'\rangle$ with  effective parameters
$\alpha'$ and $\varepsilon'=r'e^{i\pi}$. The effect of linear loss
on a mode can be described by \cite{hpyu,hyjs,rsby, hpyu1,hpyu2}
\begin{equation}\label{eq1}
b=\eta a+\sqrt{1-\eta^2}a_v,
\end{equation}
where $a_v$ is the annihilation operator of the vacuum state mode.
Defining $b_2=(b-b^\dagger)/i$, from equation \eqref{eq1}, we have
\begin{equation}\label{eq2}
\langle b\rangle=\eta \alpha=\alpha',
\end{equation}
\begin{equation}\label{eq3}
\langle \Delta b_2^2\rangle=\eta^2 e^{-r}+(1-\eta^2)=e^{-r'},
\end{equation}
where $\langle\,\rangle$ indicates an expectation values for the
squeezed state $|\eta \alpha, \varepsilon'\rangle$. When the light
pulses propagate in the optic fiber, the lost photons are
immediately lost, and the subspace of the lossy mode may be assumed
to be in vacuum at all times (Born approximation) \cite{lad}. The
pulse in the fiber can be expressed as
$|\eta\alpha,\varepsilon'\rangle\otimes|0\rangle_l$, where the
subscript $l$ denotes  lost photons. Thus, tracing over the lost
photons will not result in new decoherence, and the decoherence from
the photon loss in the channel have been embodied in the reduction
of the parameters of the pulses.

At node 2, considering the decoherence from cavity losses and
spontaneous emission during the light-matter interaction, the system
composed of the pulses and the atom in the node 1 can be described
by
\begin{eqnarray}\label{eq4}
\rho&=&|0\rangle|\eta \alpha,\varepsilon'\rangle\langle\eta
\alpha,\varepsilon'|\langle0|+|1\rangle|\eta \alpha
e^{-i\theta},\varepsilon' e^{-i2\theta}\rangle\langle\eta \alpha
e^{-i\theta},\varepsilon'
e^{-i2\theta}|\langle1|\notag\\&+&\zeta|0\rangle|\eta \alpha
,\varepsilon'\rangle\langle\eta \alpha e^{-i\theta},\varepsilon'
e^{-i2\theta}|\langle1|\notag\\&+&\zeta|1\rangle|\eta \alpha
e^{-i\theta},\varepsilon' e^{-i2\theta}\rangle\langle\eta
\alpha,\varepsilon'|\langle0|
\end{eqnarray}
where $\zeta$ describes the decoherence arising from the dispersive
light matter interaction \cite{hon} and the lost mode of photons has
been traced over.

Because at the second node, the parameters of the probe pulses
$\alpha$ and $\epsilon$ decrease to $\eta\alpha$ and $\varepsilon'$,
respectively. Then the phase shift of $\alpha$ due to the dispersive
light-matter interaction in the second cavity will be smaller than
that acquired in the first node, if other parameters of the
interaction are the same as those in the first one. This problem,
however, can be overcome by setting the atom-cavity coupling factor
$g$ at the second node to be larger than that at the first node.

We adopt a linear phase shift of $\theta$ to component $\eta\alpha$
of the probe state after it leaves node 2, which will result in a
corresponding phase shift $2\theta$ to $\varepsilon'$ (see the
following discussions). Because in the homodyne detection, only the
pulses described by $|\eta\alpha ,\varepsilon'\rangle $ ,
$|\eta\alpha e^{i\theta},\varepsilon' e^{i2\theta}\rangle$, and
$|\eta\alpha e^{-i\theta},\varepsilon'e^{-i2\theta}\rangle$ can
appear, we only need to consider the diagonal elements of the total
density matrix for the probe beams, which
 have the form
\begin{equation}\label{eq12}
\begin{split}
\rho_{dia}=&\frac{1}{4}|00\rangle\cdot \langle 00|\cdot|\eta\alpha
e^{i\theta},\varepsilon' e^{i2\theta}\rangle\langle\eta\alpha
e^{i\theta},\varepsilon' e^{i2\theta}|+\\&\frac{1}{4}|11\rangle
\langle 11|\cdot|\eta\alpha
e^{-i\theta},\varepsilon'e^{-i2\theta}\rangle\langle\eta\alpha
e^{-i\theta},\varepsilon'e^{-i2\theta}|\\&+\frac{1}{2}\rho_{en}\cdot|\eta\alpha
 ,\varepsilon'\rangle\langle\eta\alpha
 ,\varepsilon'|,
\end{split}
\end{equation}
where
\begin{equation}\label{eq13}
%\begin{split}
\rho_{en}=\frac{1}{2}(|01\rangle\langle 01|+\zeta|01\rangle\langle
10|+\zeta|10\rangle\langle 01|+|10\rangle\langle 10|).
%\end{split}
\end{equation}

The Wigner function for a squeezed state $|\alpha,\varepsilon\rangle
= |\frac{1}{2}(X_1+X_2),r\rangle$ \cite{wal} is
\begin{equation}
W(x'_1,x'_2)=\frac{2}{\pi}\exp\left[-\frac{1}{2}(x'^{2}_1e^{-2r}+x'^2_2e^{2r})\right],
\end{equation}
where $x'_i=x_i-X_i\,(i=1,2)$. Since $\theta<0.01$ rad, we may
neglect the effect of the phase shift $\pm2\theta$ of the squeeze
parameter $\varepsilon'$. Through the balanced homodyne detection of
$p$ quadrature of the squeezed pulse \cite{scu}, the success
probability of generating entanglement between two qubits in two
neighbor nodes is found to be
\begin{equation}\label{eq14}
P_s=\text{Tr}\int_{-p_c}^{p_c}\rho\,
dx_2=\frac{\text{erf}(b_0)}{2}+\frac{\text{erf}(b_1)}{4}+\frac{\text{erf}(b_{-1})}{4},
\end{equation}
where $b_s=\sqrt2(p_c+s\eta d)\exp(\varepsilon_1)$, $s=0,\pm1$,
$d=\alpha \sin\theta$,
$\text{erf}(t)=\frac{2}{\sqrt{\pi}}\int^{\,t}_0e^{-x^2}dx$, and
$p_c$ is the selection window of the homodyne measurement.
  The average fidelity
after postselection has the form
\begin{equation}\label{eq15}
\begin{split}
F&=\frac{1}{P_s}\left[\int_{-p_c}^{p_c}dx_2\langle\psi^+|\rho|\psi^+\rangle\right]\\
 &=\frac{\text{erf}(b_0)(1+\zeta)}{2\text{erf}(b_0)+\text{erf}(b_1)+\text{erf}(b_{-1})},
\end{split}
\end{equation}
with the desired entangled output state $|\psi^+\rangle=(|01\rangle
+|10\rangle)/\sqrt{2}$.

\begin{figure}
\includegraphics[scale=0.4]{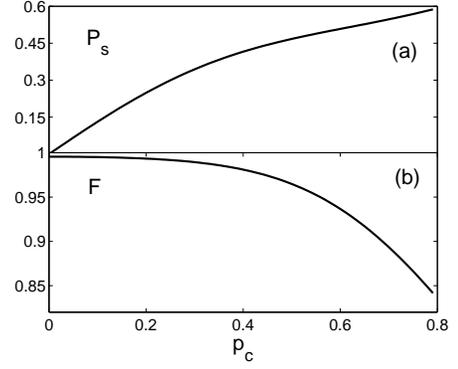}
\caption{\label{fig:2} The success probability $P_s$ of successful
generation of entanglement (a) and the average fidelity $F$ of the
obtained entanglement between two qubits in neighboring nodes (b)
 as a function of the postselection window $p_c$ for
$\eta^2=2/3$, $\alpha=150$, $\varepsilon=1.61$, $\theta=0.00867$
rad, $d=1.30$, and $\zeta=0.995$. }
\end{figure}

For optical fibers and wavelengths of visual light, the loss is
about 0.17dB/km and the transmission parameter in a length of 10 km
is $\eta^2=2/3$ \cite{loo}. 7 dB  squeezed lights have a
corresponding squeeze factor $r=1.61$. From equation \eqref{eq3}, we
have $r'=0.511$.  Setting $\alpha=150$, $\theta=0.00867$ rad,
$d=1.30$, and $\zeta=0.995$ \cite{hon}, according to equations
(\ref{eq14}, \ref{eq15}), we have the success probability
$P_s=0.344$, and the average fidelity $F=0.989$ for the selection
window $p_c=0.3$. The dependence of the success probability $P_s$
and the average fidelity $ F$ on the  selection window $P_c$ are
shown in Fig.\ref{fig:2}(a) and Fig.\ref{fig:2}(b), respectively. From Fig. \ref{fig:2}, we see that the success probability $P_s$ increases, however, due to the lack of orthogonality in the squeezed bases, the average fidelity $F$ decreases, when we choose  a larger selection window $p_c$. Thus a compromise between  the success probability $P_s$ and the average fidelity $F$ has to be made \cite{loo}.

 Now we discuss the linear phase shifter for the
squeezed state. When a squeezed state light $s$ and another very
strong reference light $t$ pass through a Kerr medium at the same
time, in which the interaction between the two beams is described by
\cite{scu}
\begin{equation}
H_{in}=\hbar \chi a^\dagger_s a_sa^\dagger_t a_t,
\end{equation}
where $\chi$ is a coupling constant dependent on the third-order
non-linear susceptibility for the optical Kerr effect, we can find

\begin{equation}
a_s(t)=e^{i\chi A_tt}a_s(0)\equiv e^{i\theta(t)}a_s(0),
\end{equation}
where $A_t=a^\dagger_t a_t$ is a constant of motion.
 Thus we have
\begin{equation}
\langle\alpha,\varepsilon|a_s(t)|\alpha,\varepsilon\rangle=e^{-i\theta(t)
}\langle\alpha,\varepsilon|a_s(0)|\alpha,\varepsilon\rangle=e^{-i\theta(t)
}\alpha,
\end{equation}
and \cite{mors}
\begin{eqnarray}
 \langle \Delta N(t)\rangle^2&=&\langle \Delta
N(0)\rangle^2=2\sinh^2r\cosh^2r\notag\\
&+&|\alpha|^2\left[e^{-2r}\cos^2(\theta-\phi)+e^{2r}\sin^2(\theta-\phi)\right].
\end{eqnarray}
where $\langle\,\rangle$ indicates an expectation value for the
squeezed state $| \alpha, \varepsilon\rangle$. By this way a linear
phase shift of $\theta$ and $2\theta$ have been applied to $\alpha$
and $\varepsilon$, respectively. Large phase shifts may be possible
in some solid-state systems such as Fullerence quantum dots which
have a very large electric dipole moment and couple strongly to the
electric field \cite{tskn}.

In conclusion, we present a scheme for distributing entanglement
between qubits at remote stations in which the entanglement between
qubits is generated by using a bright light pulse in the squeezed
state. Together with a high success probability, the average
fidelity of the abtained entanglement between qubits at the
neighboring stations which are 10 km apart from each other may reach
near unity without entanglement purification. This scheme, combined
with protocols for entanglement purification and swapping such as
the protocol suggested by \cite{chil}, may be a promising candidate
for the quantum repeaters. It may considerably increase the rates of
quantum communications.

 {\it Acknowledgments} This work was supported by the State Key Programs for Basic Research
of China (2005CB623605 and 2006CB921803), and by National Foundation of Natural Science in China
Grant Nos. 10474033 and 60676056.


\begin{references}
\bibitem{Giulio}G. Chiribella, L. Maccone, and P. Perinotti,  Phys. Rev. Lett. {\bf 98}, 120501
(2007).
\bibitem{dur} W. D$\ddot{u}$r {\it et al.}, Phys. Rev. A {\bf 59}, 169
(1999).
\bibitem{bri} H.J. Briegel, W. Dur, J.I. Cirac, and P. Zoller,  Phys. Rev. Lett. {\bf 81}, 5932
(1998).
\bibitem{ekert}A. Ekert,  Phys. Rev. Lett. {\bf 67}, 661 (1991)
\bibitem{bra}G. Brassard, N. Lutkenhaus, T. Mor, and B.C. Sanders, Phys. Rev.
Lett. {\bf 85}, 1330 (2000)
\bibitem{ben}C. Bennett {\it et al.}, Phys. Rev. Lett. {\bf 76}, 722 (1996).
\bibitem{deu}D. Deutsch {\it et al.}, Phys. Rev. Lett. {\bf 77}, 2818 (1996).
\bibitem{cir}J.I. Cirac, P. Zoller, H.J. Kimble, and H. Mabuchi, Phys. Rev. Lett. {\bf 78},
3221 (1997).
\bibitem{yao}W. Yao, R.-B. Liu, and L.J. Sham,  Phys. Rev. Lett. {\bf 95}, 030504 (2005).
\bibitem{kraus}B. Kraus, and J.I. Cirac,  Phys. Rev. Lett. {\bf 92}, 013602 (2004).
\bibitem{duan}L.-M. Duan, M.D. Lukin, J.I. Cirac, and P. Zoller, Nature {\bf 414},413 (2001).
\bibitem{chi}L. Childress, J.M. Taylor, A.S. S$\phi$rensen, and M.D. Lukin,  Phys. Rev. A
{\bf 72}, 52330 (2005).
\bibitem{chil}L. Childress, J.M. Taylor, A.S. S$\phi$rensen, and M.D. Lukin,  Phys. Rev. Lett.
{\bf 96}, 070504 (2006).
\bibitem{wak}E. Waks, and J. Vuckovic, Phys. Rev. Lett. {\bf 96}, 153601 (2006).
\bibitem{duanl}L.-M. Duan, and H.J. Kimble, Phys. Rev. Lett. {\bf 92}, 127902 (2004).
\bibitem{loo}P. van Loock, T.D. Ladd, K. Sanaka, F. Yamaguchi, K. Nemoto, W.J. Munro,
 and Y. Yamamoto, Phys. Rev. Lett. {\bf 96}, 240501 (2006).
\bibitem{lad}T.D. Ladd, P. van Loock, K. Nemoto, W.J. Munro, and Y. Yamamoto,  New J. Phys.
{\bf 8}, 184 (2006).
\bibitem{wsmk} W. Son, M.S. Kim, J. Lee, and D. Ahn, J. Mod. Opt. {\bf 49}, 1739 (2002).
\bibitem{mpws} M. Paternostro, W. Son, and M.S. Kim, Phys. Rev. Lett. {\bf 92}, 197901 (2004).
\bibitem{bkjc} B. Kraus and J.I. Cirac, Phys. Rev. Lett. {\bf 92}, 013602 (2004).



\bibitem{pklr}P.K. Lam, T.C. Ralph, B.C. Buchler, D.E. McClelland, H-A. Bachor, and J. Gao, J. Opt. B {\bf 1},469 (1999).

\bibitem{guz}R. Guzm\'{a}n, J.C. Retamal, E. Solano, and N. Zagury,  Phys. Rev. Lett. {\bf 96},
010502 (2006).
\bibitem{kni}E. Knill, R. Laflamme, and G.J. Milburn, Nature {\bf 409}, 46 (2001).
\bibitem{nem}K. Nemoto, and W.J. Munro, Phys. Rev. Lett. {\bf 93}, 250502 (2004).
\bibitem{abra} A.S. Bracker {\it et al.}, Phys. Rev. Lett. {\bf 94}, 047402 (2005).
\bibitem{str}S. Strauf {\it et al.}, Phys. Rev. Lett. {\bf 89}, 177403 (2002).
\bibitem{fu}K.-M.C. Fu {\it et al.}, Phys. Rev. B {\bf 69}, 125306 (2004).
\bibitem{wal}D.F. Wall, and G.J. Milburn, {\it Quantum Optics}, Springer-Verlag, Berlin, 1995.
\bibitem{hon}F.-Y. Hong, and S.-J. Xiong arXiv:quant-ph/0804.0720v2.
\bibitem{hpyu},H.P. Yuen, arXiv:quant-ph/0109054v1.
\bibitem{hyjs} H.P. Yuan and J.H. Shapiro, IEEE Trans. Inform. Theory. {\bf 24}, 657 (1978).
\bibitem{rsby} R.E. Slusher and B. Yurke, J. Lightwave Technol. {\bf 8}, 466 (1990).
\bibitem{hpyu1}H.P. Yuen, " Quantum Communication, Quantum Measurement, TCS and QND" in {\it Quantum Optics, Experimental Gravity, and Measurement Theory}, ed. P. Meystre and M.O.Scully, Plenum, pp.249-268, 1983.
\bibitem{hpyu2}H.P. Yuen, "Nonclassical Light" in {\it Photons and Quantum Fluctuations}, ed. E.R. Pike and H. Walther, Adam Hilger, pp.1-9, 1988.
\bibitem{scu}M.O. Scully, and M.S. Zubairy, {\it Quantum Optics}, Cambridge, 1997.
\bibitem{mors} M. Orszag 2000 {\it Quantum Optics} (Berlin: Springer-Verlag).
\bibitem{tskn} T.P. Spiller, K. Nemoto, S.L. Braunstein, W.J. Munro, P. van Loock, and G.J. Milburn,  New J. Phys.
{\bf 8}, 30 (2006).

\end{references}
\end{document}